# Using Wireless Sensor Networks to Narrow the Gap between Low-Level Information and Context-Awareness

Ioan Raicu[†][*], Owen Richter[‡], Loren Schwiebert[†], Sherali Zeadally[†]

[†]Computer Science Department
Wayne State University
Detroit, MI 48202, USA
{iraicu, loren, zeadally}@cs.wayne.edu

[‡]Accenture Technology Labs
Accenture
Palo Alto, CA 94304, USA
owen.richter@accenture.com

**Abstract**

Wireless sensor networks are finally becoming a reality. In this paper, we present a scalable architecture for using wireless sensor networks in combination with wireless Ethernet networks to provide a complete end-to-end solution to narrow the gap between the low-level information and context awareness. We developed and implemented a complete proximity detector in order to give a wearable computer, such as a PDA, location context. Since location is only one element of context-awareness, we pursued utilizing photo sensors and temperature sensors in learning as much as possible about the environment. We used the TinyOS RF Motes as our test bed WSN (Wireless Sensor Network), 802.11 compatible hardware as our wireless Ethernet network, and conventional PCs and wired 802.3 networks to build the upper levels of the architecture.

## 1 INTRODUCTION

Over the last half a century, computers have exponentially increased in processing power and at the same time decreased in both size and price. These rapid advancements led to a very fast market in which computers would participate in more and more of our society's daily activities. In recent years, one such revolution has been taking place, where computers are becoming so small and so cheap, that single-purpose computers with embedded sensors are almost practical from both economical and theoretical points of view. The next logical step in combining wireless sensor networks and context aware wearable computers is to build an infrastructure, and make the technology easy enough to be implemented by non-specialists.

In order to better understand the problem this paper is concentrating on, both low-level information and context-awareness should be clearly defined. The low-level information consists of the raw sensor readings data, such as temperature, light intensity, radio signal strength, etc. However, humans do not relate well to raw numbers read from the sensor, and therefore at the very least, the system should translate the raw number to a known scale, such as Fahrenheit for temperature. This is better, but it is still not enough, because having a temperature in Fahrenheit still does not indicate what it means, and therefore thresholds must be set to identify what cold, moderate, and hot mean in a certain context. Then, if the computer knows that it is cold in the room, it would turn up the heat. Therefore, context-awareness as we will be using it throughout this paper is the condition in human understandable terms in which a computer or application is operating; of course, the computer or application should adapt its operating requirements depending on the context. Thus, a system that behaves in such a fashion is termed as being context-aware.

Our process of converting low level information to high level information is divided into three parts. The first is the data acquisition, which until recently has been very difficult because of the limited application of wired sensors. In our case, we are using wireless RF sensors which run the TinyOS [3]. A future method could potentially use "Smart Dust". "Smart Dust" is a Micro-Electro Mechanical Systems (MEMS) which are essentially micron scale wireless sensors [6]. In order to connect the acquisition stage with the processing stage, we use two technologies already in wide use and acceptance, IEEE 802.3 [4] networks and IEEE 802.11 [5] networks. 802.3 networks have a wired physical medium of communication, such as copper or optical fiber. On the other hand, 802.11 networks have a wireless RF communication medium, but use the same Ethernet frames as their wired counterparts, which allows them to interoperate easily. The second part is to preprocess all the data and store the results (state information about each node) in a local database. In order to link the second phase with the third one, the same communication technologies are used, both wired and wireless Ethernet networks. The third stage is to retrieve the state information about certain nodes, which should give information about the context of the environment such as location, temperature, light levels, etc.

To some extent, the systems "MetaPark" in [1] and "GUIDE" in [2] have both addressed some of the issues that we discussed thus far. The difference between our proposed work and what has already been done is the generalization of making a system that is scalable and universally usable. This is a very powerful statement, however with keeping the end result in close focus, the right decisions can be made in these early stages of this new technology. Our ultimate goal is to make wearable computers context-aware everywhere and give them the capability of adapting their behavior depending on the context in which a certain application will be operated.

In this paper, we exemplify several applications utilizing context-awareness (Section 2), we then propose a general architecture to obtain context-awareness from low-level information (Section 3), explain the proximity detection algorithm in detail (Section 3.2), describe our preliminary experimental results obtained by implementing a pilot system based on the proposed design (Section 4), and finally conclude (Section 5).

## 2 MOTIVATION AND GOALS

In terms of localization techniques, one might ask the question regarding why GPS (Global Positioning System) is not sufficient in handling some of the problems stated in this paper, especially that there already is a global infrastructure deployed, orbital satellites. The answer lies in having both indoor and outdoor localization at the same time. GPS only offers this when it has a clear view of the sky, mostly in outdoor situations. We are trying to expand the usability of a localization tool that would use GPS for outdoor localization and at the same time it would use a wireless sensor network to achieve the same thing indoors. By combining these two technologies, we can essentially create an end-to-end proximity detector that would work both outdoors and indoors.

Localization is the most prominent application that this architecture brings to mobile computing. Having a wearable computer know its position at all times is a huge leap forward in making the computer interact depending on the location. For example, if a person is at his desk, he might want to receive certain messages while if he is at a restaurant at lunch, he might want to receive a different set of messages, or perhaps if he is on vacation on the beach he might want to receive yet another different set.

Obviously, this ties in directly to getting information about the location you are at, whether it is about the next meeting for a person at work, to what the movie times and summaries are for a person at the movie theater, to what the menu is at the restaurant, and so on. All these applications would automatically help the user without asking for much interaction. For this idea to become a reality, beacons (wireless nodes) must be placed throughout the entire coverage of the infrastructure and programmed to emit a unique ID. The user that would be in proximity of the beacons would then receive the signal on the wearable computer, such as a PDA, via the integrated wireless sensor node, after which it would query the local server (similar to the foreign agent in mobile IP [7]) via the 802.11 wireless network, for information regarding the beacon that it heard. The local server would respond back with either the information itself or perhaps a URL to a web site with the corresponding information. We plan on pursuing this idea by implementing such a pilot system in the near future.

Yet another example is utilizing the sensor information to figure out the context in which the computer is situated. For example, if there are many people in a room, most likely the temperature will increase and the volume level is much higher than if there are fewer people in the room. This could immediately prompt the ventilation system in the room to increase the air flow and decrease the temperature. This could also prompt a wearable computer to have louder audible alerts in order to get the attention of the user. Photo sensors can be utilized in realizing the lighting conditions and adjusting the brightness and contrast on the display screen automatically. Vibration sensors could realize if the user is sitting at a desk or traveling in a car and therefore modify the handwriting recognition characteristics due to the different style of writing.

There are endless examples where wireless sensor networks could be used to realize context-awareness in wearable computers; however, none of them can be realized without a very much needed scalable infrastructure. The significance of our work lies in implementing a system that can give any general purpose computer or application context-awareness.

## 3 PROPOSED ARCHITECTURE

We are going to concentrate most of our efforts on building a system with off-the-shelf components, which might include production and experimental hardware and software.

The hardware and software is comprised of devices at 4 different levels. At the bottom level, we have the RENE RF motes (MMS and FMS in Fig. 1) with the TinyOS that takes samples from their sensors, or perhaps the active badge utilizing the infrared spectrum, or the GPS, or any other sensor device. This level represents the acquisition level introduced in Section 1. At the next level, we have a wearable computer (MWC in Fig. 1)

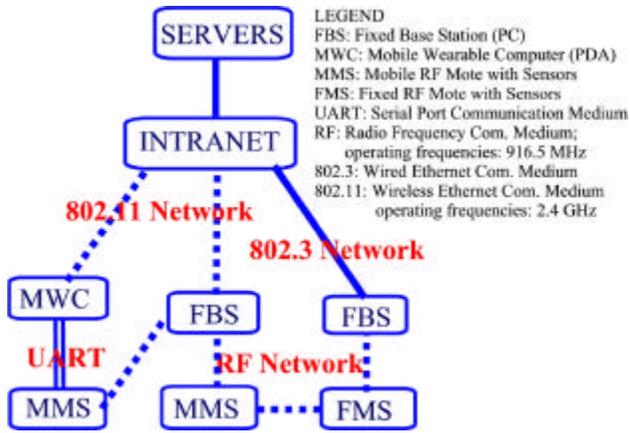

Figure 1: The proposed architecture at the BuL and WSN Levels; lines in general represent communication mediums; double solid lines represent the UART; dotted lines represent wireless communication, whether it is via 802.11 standard, or the WSN; and solid lines represent wired physical medium such as copper or optical fiber and uses 802.3 standard

such as a PDA, which might or might not be equipped with either a wireless sensor for communication with the sensor network, or perhaps an 802.11 wireless Ethernet network card to communicate with the intranet. Next there would be base stations (FBS in Fig. 1), which will most likely be workstations running standard operating systems such as Microsoft Windows and also have wireless capabilities to communicate with the sensor network and also possibly a wired (802.3)/wireless (802.11) connection to the local intranet. There is one more upper level, consisting of one or more servers that communicate through the wired/wireless intranet with all the base stations and maintains a consistent database with state information of the wireless sensor network. This state information could be comprised of numerous items, such as location information, online/offline status, sensor information, etc. Levels 2, 3, and 4 are part of the processing level that takes the input from the sensors, analyzes it and writes the state information into a database. The last stage is the retrieval process in which a MWC (Mobile Wearable Computer) interrogates the database and concludes a context in which it must operate. Obviously, we have only addressed the problem on a local intranet which is most likely limited to a physical building, however, the same system can be further expanded by adding more levels to give the system a universal architecture in which every wireless sensor node becomes an online object. The expanded architecture is briefly presented in Section 5 and will be the subject of a future paper.

The entire global proposed system is described using three diagrams shown in Fig. 1, 2 and 7. In this paper, we concentrate our discussion on the parts of the system illustrated by Fig. 1 and 2, while Fig. 7 illustrates the universal architecture and thus, encompasses both the Building Level (BuL) and the Wireless Sensor Network (WSN) from Fig 1 and 2, respectively.

To better understand how the wireless sensor network nodes communicate with each other, consider Figure 2 below. In this diagram, we present only the WSN that is presented in both Fig. 1 and 7, but at the level of communication messages.

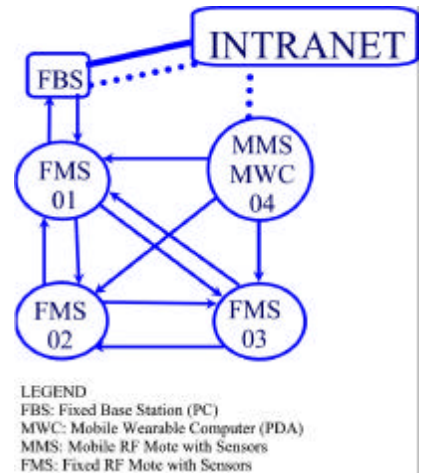

Figure 2. Sample communication in WSN

In Fig. 2, FMS 01 is a base station connected to a PC; this can be viewed as the gateway between the wireless sensor network and the wired/wireless Ethernet network and both comprise the FBS. FMS 02 and 03 are hybrid base stations due to the fact that they are stand-alone wireless sensor nodes which communicate with all other nodes within range and communicate useful information to FBS, such as proximity detection, photo and sensor information, etc. Mote 04 is a mobile sensor node which could be either solo on a badge, or perhaps work in conjunction with a wearable computer (MWC). The nodes are self-configurable in terms of finding their base stations and routing useful information to the right node; the routing is all handled within the TinyOS.

### 3.1 Sensor Hardware and Software

We used the RENE RF Mote which runs the TinyOS, a very small footprint event driven operating system designed for computers with limited processing power and battery life. While still under development and very early in its life cycle, the technology seems very promising due to the fact that the OS and its applications are written in a C - like programming language that already has a very broad base of users. A mote consists of a mote motherboard with a 4 MHz Atmel processor, 512 bytes of SRAM, 8KB of flash memory, a 916.5 MHz RF transceiver, and an antenna.

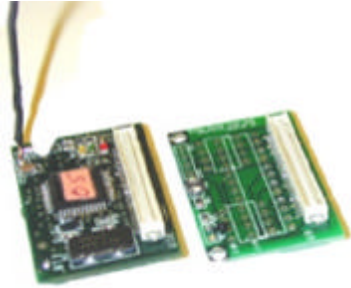

Figure 3: RENE RF Mote

An optional sensor board that plugs into the mote motherboard and includes both a photo and temperature sensor can be used as well. Our current implementation allows up to five more analog sensors in addition to the existing photo and temperature sensors.

At the moment, it requires technical specialists in order to successfully integrate a system so complex, but our hopes are that as this technology matures and the infrastructure grows, it will be as easy as turning on a wireless sensor node to be able to integrate it to the existing network.

### 3.2 Proximity Detection Algorithm

Our implementation is based on the proximity detection algorithm described below and makes use of the following concepts. A *message* is a set of information from a wireless sensor node. The typical information consists of the source of the message, destination, signal strength (which is directly proportional with the distance between the source and destination of the message), and the timestamp. The *timestamp* is a real time value associated with the reception of the message; notice that logical timestamps could also be used to simplify the algorithm. The *memory* is represented by a 3-dimensional array which can be visualized in Fig. 4. Axes X and Y correspond to the destination and source of a message respectively; the data element of the array consists of the RF signal strength; finally, the Z axis represents the timeline and hence is the history of the previous states. A *state change* occurs when the location of the mobile RF mote changed with a good confidence. A *good confidence* is defined by step 3 if criteria a, b, and c are all satisfied. The *last known location* is derived by comparing the signal strengths of all the last set of messages; realize that there is no confidence that the location is indeed accurate. The *last confirmed location* is derived by step 4 below which produces a good confidence.

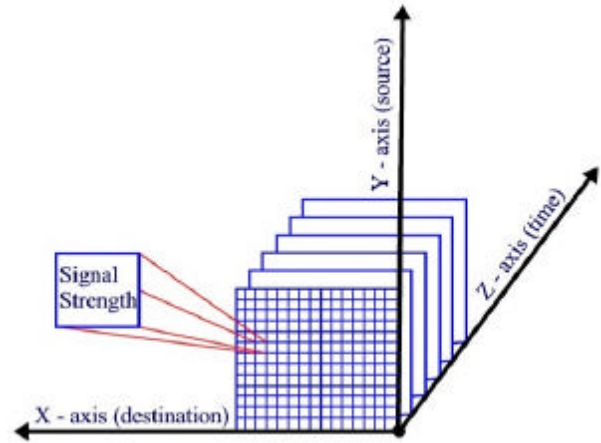

Figure 4. The memory layout: used to derive the confidence that a state has changed.

Our algorithm is broken down into five steps. All fice steps must be executed sequentially forever, or until the user decides to terminate the process.

1. Receive message and timestamp it
2. Update memory which includes adding new entries and deleting expired entries (only retain the last 6 seconds of information)
3. Search memory for a state change defined according to the following criteria
   a. Is the last known location different than the last confirmed location?
   b. Are there at least 3 messages in the history announcing the new location?
   c. Is the average of the signal strengths of the last 3 messages greater than any other location for the last 3 messages?
4. If a, b, and c are all true in Step 4, then replace the last confirmed location with the last know location
5. Display the last confirmed location

It is evident that both Steps 2 and 3 will heavily influence the performance of the algorithm. Increasing the memory of the algorithm and requiring stricter tests in searching for state changes might make the algorithm more suitable for environments in which much interference is present. Decreasing the memory and the strictness of the same tests might prove to produce faster algorithms, but will decrease the accuracy of the system.

## 4 EXPERIMENTAL RESULTS

In our experiments, we started the problem of proximity detection using the wireless RF motes from Fig. 3. We used the approach of having fixed beacons throughout an area, a base station which acted as the gateway to the Intranet to update the central database, and a badge with a RF mote identifying the user. According to [8], approximate localization is possible using RF signal strength readings and triangulation.

We implemented our system using those ideas and proved that proximity can be established by calculating signal strengths between the various wireless nodes and aggregating all the information at the base stations, which then updates a database on a local server with the new state information of all the nodes in range. With the use of history information regarding the state of each node, the location can be approximated with a resolution correlated only with the node density. In the event of a badge system in which each badge would have a wireless node, the closest base station would calculate the proximity of the badge in relation to the other nodes. In the event of a wearable computer, the PDA would use its 802.11 network resources in retrieving necessary information regarding its location.

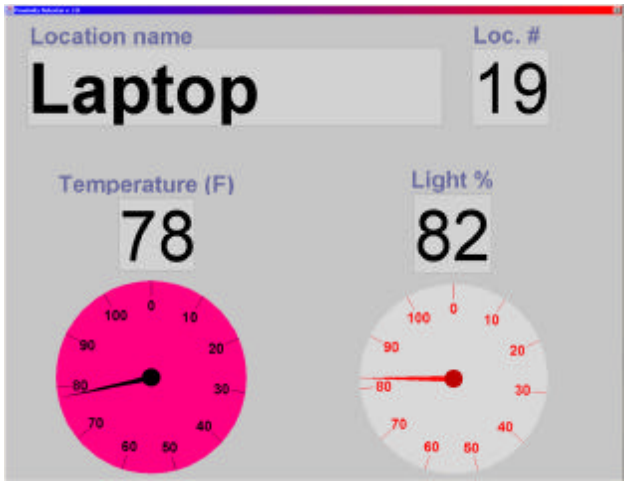

Figure 5: Current implementation displaying location information and sensor readings already converted to human readable numbers, such as Fahrenheit temperature and percentage of light present

The current user interface is meant to convey the location's ID and description, and the sensor information at the corresponding location. Fig. 5 shows a screen shot of the interface. Our demo was implemented on a laptop computer running Windows 2000, but the same interface can easily be ported to a PDA.

In order to convey the performance of the described architecture, we conducted a set of experiments in an open 15 by 20 ft. rectangular room. In positioning the sensor nodes throughout the room, we used the schema from Fig. 2: the base station was positioned in one of the corners, while three other FMS (fixed RF motes) were placed in the other 3 corners of the room. When the MMS (mobile RF mote) walks in the room, its presence will be detected and its position will be transmitted to the base station. When the MMS walks around the room, the base station will recalculate the MMS's position at every message in terms of which FMS it was closest to. The results in Fig. 6 were achieved by repeatedly walking around the room and measuring the accuracy of the state changes, the average time to detect the state change, and the maximum distance that the algorithm realized the state change. The maximum distance we obtained can easily be smaller or greater depending with what rate of speed the MMS is moving.

| Accuracy to Detect State Changes | 100% |
|---|---|
| Average Time to Detect State Change | 3 seconds |
| Maximum Distance to Detect State Change | 6 feet |

Figure 6: Performance metrics and results for the proposed algorithm in Section 3.2

The results are also heavily dependent on the parameters set within our proximity detector algorithm described in Section 3.2. Increasing the memory of the algorithm and requiring stricter tests in searching for state changes might make the algorithm more suitable for environments in which much interference is present, but it might increase the average time taken to detect a state change. Decreasing the memory and the strictness of the tests might prove to produce faster algorithms, but will decrease the accuracy of the system; notice that if only the last message received is taken into consideration, and the mobile RF mote is half way between two fixed RF motes, the state of the mobile RF mote could be changing at each message producing an output that makes no sense. The algorithm's parameters we chose, such as having a six second memory and taking the average of the last three messages, gave the algorithm a good enough confidence when a state change was detected.

## 5 CONCLUSIONS AND FUTURE WORK

In this paper, we discussed a possible architecture for making context-aware applications a reality. As a quick reiteration, we used the TinyOS RF motes and 802.11 compatible hardware, and off-the-shelf workstations to realize the test-bed. We utilized proximity detection indoors, photo sensors, and temperature sensors to give wearable computer context information about its environment. Our system is scalable due to its hierarchical structure. Our preliminary results at the lower levels of the hierarchy seem very promising.

As future work, we plan on integration of the RF mote into a PDA in order to allow more mobility for the user. This would then allow the architecture to be tested more thoroughly as we deploy the wireless sensors throughout the building.

A further step would be to have the hierarchical system in Fig. 7 that is location-based and would allow a unified infrastructure very similar to what the Internet has evolved to today. Placing servers at all these various levels would essentially place every wireless sensor node online, and have it identified according to its location and possibly function.

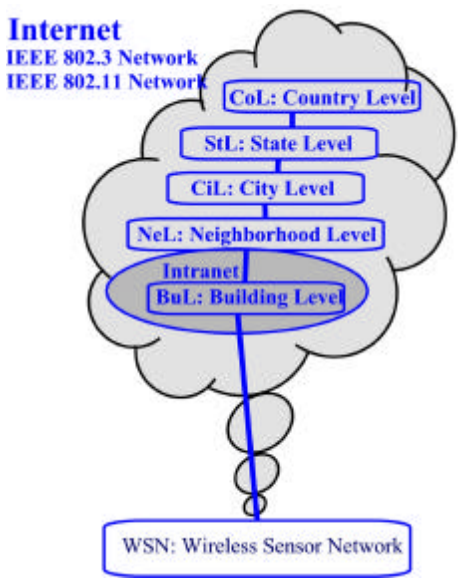

Figure 7: Possible future hierarchical system to support wireless sensor nodes in having an on-line presence

The proposed system is scalable due to the fact that each level has only to know about itself, the level above it, and the one below it. Therefore the overhead of having billions of nodes becomes very distributed. Moreover, each level is responsible for the level adjacent to itself. This approach has its similarities with the class-based approach of internet addressing, in which each network administers its own addresses. Once such a system is in place, an object such as a wireless sensor node could be identified by its GUA (Globally Unique Address): CoL.StL.CiL.NeL.BuL.LocalUniqueID (the acronyms in the GUA are defined in Fig. 7)

We hope that the ideas and experimentation presented in the paper will have a great impact on this new and emerging technology and will help others on developing a global infrastructure to make context-awareness a part of tomorrow's computers.

## 6 REFERENCES

[1] N. Dyer, J. Bowskill, "Ubiquitous Communications and Media: Steps Toward a Wearable Learning Tool", British Computer Society Conference on 'Digital Media Futures', Bradford, UK, 1999.

[2] K. Cheverst, G. Smith, K. Mitchell, N. Davies, "Exploiting Context to Support Social Awareness and Social Navigation", Proceedings of the Workshop on 'Awareness and the WWW' at CSCW '00, Philadelphia, 2000.

[3] J. Hill, R. Szewczyk, A. Woo, S. Hollar, D. Culler, K. Pister, "System architecture directions for network sensors", ASPLOS 2000.

[4] A. S. Tanenbaum, Computer Networks, Third Edition, Prentice Hall Inc., 1996, pp. 276-287.

[5] G Coulouris, J. Dollimore, T. Kindberg, Distributed Systems, Concept and Design, Third Edition, Addison-Wesley Publishers Limited, 2001, pp. 116-119.

[6] J. Kahn, R. H. Katz, K. Pister, "Emerging Challenges: Mobile Networking for 'Smart Dust'." Journal of Communications and Networks, vol. 2, no. 3. pp. 188-196, September 2000.

[7] A. S. Tanenbaum, Computer Networks, Third Edition, Prentice Hall Inc., 1996, pp. 367-370.

[8] L. Doherty, K. S. J. Pister, L. E. Ghaoui, "Convex Position Estimation in Wireless Sensor Networks" Infocom 2001, Anchorage, AK, 2001.